\documentclass[12pt]{article}
\usepackage{amssymb,amsfonts}
\usepackage[top=2.5cm, bottom=2.75cm, left=2.5cm, right=2.5cm]{geometry}

\newcommand{\be}{\begin{equation}}
\newcommand{\ee}{\end{equation}}
\newcommand{\beqs}{\begin{eqnarray}}
\newcommand{\eeqs}{\end{eqnarray}}

\newcommand{\dpl}{{\dot+}}
\newcommand{\dm}{{\dot-}}
\newcommand{\nt}{{\tilde n}}
\renewcommand{\a}{\alpha}
\renewcommand{\b}{\beta}

\newcommand{\pa}{\partial}
\newcommand{\na}{\nabla}
\newcommand{\g}{\gamma}

\newcommand{\e}{\epsilon}
\newcommand{\z}{\zeta}
\newcommand{\zt}{{\tilde\zeta}}

\renewcommand{\l}{\lambda}

\newcommand{\pis}{{\pi\kern-1.28ex /}}
\newcommand{\ds}{{\partial\kern-1.28ex /}}
\newcommand{\ns}{{n\kern-1.23ex /}}
\newcommand{\nts}{{{\tilde n}\kern-1.23ex /}}

\renewcommand{\o}{\omega}
\newcommand{\q}{\theta}

\begin{document}
\setcounter{page}{0}
\thispagestyle{empty}
\begin{titlepage}
\begin{center}
\hfill YITP--SB--06--14  \\
\hfill ITFA--2006--20  \\
\hfill UUITP--08/06 \\
\hfill HIP-2006-23/TH\\
\vskip 25mm

{\Large {\bf SIM(2) and Superspace}}

\vskip 8mm
{\bf Ulf Lindstr\" om$^{a,b},$ and Martin Ro\v{c}ek$^{c,d}$}
\bigskip

{\small\it
$^a$Department of Theoretical Physics 
Uppsala University, \\ Box 803, SE-751 08 Uppsala, Sweden \\[4mm]
$^b$HIP-Helsinki Institute of Physics, University of Helsinki,\\
P.O. Box 64 FIN-00014  Suomi-Finland\\[4mm]
$^{c}$C.N. Yang Institute for Theoretical Physics\\
SUNY, Stony Brook, NY 11794-3840, USA\\[4mm]
$^{d}$Institute for Theoretical Physics University of Amsterdam, \\
Valckenierstraat 65, 1018 XE Amsterdam, The  Netherlands
}\vskip 4mm
{\tt ulf.lindstrom@teorfys.uu.se}\\
{\tt rocek@insti.physics.sunysb.edu}
\vskip 6mm
\end{center}
\vskip .2in
\begin{center} {\bf ABSTRACT } \end{center}
\begin{quotation}\noindent
In this brief note we give a superspace 
description of the supersymmetric
nonlocal Lorentz noninvariant actions 
recently proposed by Cohen and Freedman. This leads us
to discover similar terms for gauge fields.
\end{quotation}
\vfill
\flushleft{\today}
\end{titlepage}
\eject
\small
\vspace{1cm}
\normalsize

\section{Introduction}
\setcounter{equation}{0}
Recently, Cohen and Glashow proposed that certain nonlocal terms
that preserve just a SIM(2) subgroup of the Lorentz group
may account for neutrino masses without the need to introduce
new particles \cite{cg}. This proposal was a follow-up to the
curious observation that there does not seem to be experimental
evidence to rule out dynamics that is governed by this solvable subgroup 
of the Lorentz group \cite{cg1}.

Subsequently, these neutrino mass terms were supersymmetrized \cite{cf}.
In this brief note, we give a superspace description of these terms. The
resulting superspace action is compact and easy to understand; as
we give a superspace measure for these terms, it is easy to write down
many generalizations. In particular, we find Majorana like mass
terms for chiral superfield fermions as well as mass terms for
gauginos; the latter induce nonlocal mass terms for the
gauge fields, which become local in a lightcone gauge;
though superficially one might think that these terms
provide an alternative to the Higgs mechanism, we argue that this 
is not the case\footnote{Cohen and Glashow have independently considered
these terms \cite{fpc}.}.

We note that the SIM(2)-symmetric
but Lorentz-noninvariant terms have 
some resemblance to boundary terms; it would
be very interesting if they could be given such an 
interpretation.\footnote{K. Skenderis has pointed out that 
the intersection of a 3-brane and a 7-brane
can be interpreted as giving rise to an effective action
with a bulk and a boundary term that both fill
the world-volume of the 3-brane.}

\section{Four component notation}
\setcounter{equation}{0}
The SIM(2) group is the subgroup of the 
four dimensional Lorentz group that preserves
a fixed null vector up to rescalings; a 
nice summary of all the relevant facts
that we use can be found in \cite{cf}.
In this section, we present two forms of the SIM-invariant dynamics in superspace.
The first is formulated in terms of a constant null vector $n$; the second is 
formulated in terms of a constant spinor $\z$.
\subsection{Null vector formalism}
We begin with a few preliminaries.
In addition to the given null vector $n$, 
we may choose a second null vector $\nt$ that obeys $n\cdot\nt=1$. 
Then, since
\be
n\cdot\nt=1~\Rightarrow~ \{\ns,\nts\}=2~~,~
\left( \frac{\ns\nts}2\right)^{\!\!2}=\left( \frac{\ns\nts}2\right)~,~~etc.,
\ee
we can split any spinor $\psi$ into 
$\frac12\ns\nts\psi+\frac12\nts\ns\psi$ projections; furthermore,
the constraint $\ns\e=0$ has the obvious solution $\e=\ns\eta$ for an arbitrary
spinor $\eta$.  

We keep the Lorentz-invariant part of the
action for the chiral superfield $Z$ in full superspace:
\be
S_{full}=\int d^4\q\,Z\bar Z+\left(\int d^2\q \, W(Z) + c.c.\right)~;
\ee
To write the Lorentz-breaking term, 
we split the spinor derivative $D$ into the piece that we keep,
$d\equiv\frac12\nts\ns D$ and the piece that we expand in: 
$q\equiv\frac12\ns\nts D$. 
The SIM-superspace algebra of the spinor derivatives is \cite{cf}
\be
[\bar\e_1d\,,\bar d\e_2]=2(\bar\e_1\nts\e_2) (n\cdot \pa) 
\ee
for arbitrary constant spinors $\e_1,\e_2$.
We relate the full superspace chiral superfields
$Z,\bar Z$ obeying
\be
D_RZ=0~,~D_L\bar Z=0
\ee
to superfields that we use in SIM-superspace, 
which are the covariant spinor 
projections\footnote{We use the same
name for the full superfields and their leading
SIM-superspace projections. The SIM-superfields
are defined by the standard method of covariant
projection; equivalently, we may write the explicit
$\q$-expansion:
$Z_{full}=Z_{SIM}+\frac12\bar\q\ns\nts\psi_L
-\bar\q\nts\g^\mu\ns\q_L\pa_\mu Z_{SIM}$.} 
of $Z,\bar Z$ that are independent 
of $\frac12\nts\ns\q$, the conjugate of $q$: $Z, \bar Z$, 
and new spinor superfields $\psi_L,\psi_R$
defined by:
\be
\psi_L=q Z~,~\psi_R=q\bar Z~.
\ee
The SIM-superspace constraints that these superfields obey are
\be\label{con}
d_R Z=0~,~d_L\bar Z=0~,~
d_R\psi_L=\{d_R\,,q_L\}Z~,~d_L\psi_R=\{d_L\,,q_R\}\bar Z~,
\ee
where the anticommutators $\{d_R\,,q_L\}$ and $\{d_L\,,q_R\}$ 
are certain projections of $i\pa$ which we give explicitly 
in two component form below.
The extra SIM(2) invariant but non-Lorentz invariant term 
is\footnote{The full superspace measure can be related
to the SIM-superspace measure by 
$\int d^4\q\propto \int (\bar d\ns d_L)( \bar q \nts q_R)$.}
\be\label{4boun}
S_{SIM}=m^2\int (\bar d\ns d_L) \, Z\frac1{n\cdot\pa}\bar Z~.
\ee
Neither the SIM-superspace measure nor the SIM-superspace 
Lagrange density is SIM(2) invariant, but they transform so
as to ensure that the action $S_{SIM}$, which is
homogenous in the null vector $n$, is invariant.
Note that the superfield $\psi$ does not enter $S_{SIM}$.
We work out the component form of the action below. 
We define component fields $Z, \chi_L=d_LZ, \psi_L, 
F=d_L\psi_L$ and the conjugates 
$\bar Z, \chi_R=d_R\bar Z, \psi_R, 
\bar F=d_R\psi_R$, and push in the $d$'s
to find the action. We also need to use 
the constraints (\ref{con}). A useful identity is:
\be
\bar d\g^\mu d=\nt^\mu\bar D\ns D~
\Rightarrow \bar d\ns d = \bar D\ns D~.
\ee
\subsection{Spinor formalism}
The null condition $n\cdot n =0$ is solved in terms of a single
commuting Majorana spinor $\z$:
\be
n^\mu=\bar\z\g^\mu\z~.
\ee
Clearly, $\ns\z=0$; hence the SIM-supersymmetry transformations
are generated by a spinor 
\be
\e=(a+ib\g_5)\z~,
\ee
where $a,b$ are real {\it anti}commuting scalar parameters. Similarly,
we write $\nt$ in terms of a second commuting Majorana spinor $\zt$:
\be
\nt^\mu=\bar\zt\g^\mu\zt~,~ \bar\z\zt=2~.
\ee
Then we write:
\be
d_L\equiv\zt_L(\bar\z D_L)~,~d_R\equiv\zt_R(\bar\z D_R)~,~
q_L\equiv\z_L(\bar\zt D_L)~,~q_R\equiv\z_R(\bar\zt D_R)~,
\ee
and the SIM(2) invariant action becomes
\be\label{spin}
S_{SIM}=m^2\int (\bar\z D_L)( \bar\z D_R)
\left( Z\frac1{n\cdot\pa}\bar Z\right).
\ee
\section{Two component notation}
\setcounter{equation}{0}
This can all be made more transparent in two component notation.
The only nonvanishing components of $n$ and $\nt$ can be chosen to be
$n^{+\dpl}$ and $\nt_{+\dpl}$, or equivalently, the only nonvanishing
components of the commuting spinors $\z$ and $\zt$ can be chosen to be 
$\z^+,\bar\z^\dpl$ and $\zt_+,\bar\zt_\dpl$.
Then the spinor derivatives\footnote{Two component spinors 
$D_\a$, $\bar D_{\dot\a}$ correspond to four component spinors 
$D_L$, $D_R$, respectively.} $d,q$ are simply:
\be\label{dqdef}
d_+=D_+~,~q_-=D_-~,~ \bar d_\dpl=\bar D_\dpl~,~
\bar q_\dm=\bar D_\dm~,
\ee
where the undotted indices correspond to left-handed spinors and
the dotted indices correspond to right-handed spinors (up to a convention).
The algebra of the spinor derivatives is also very simple
\be
\{d_+\,,\bar d_\dpl\}=i\pa_{+\dpl}~,
\ee
where $\pa_{+\dpl}=n\cdot\pa$, $\pa_{-\dm}=\nt\cdot\pa$, {\it etc.}
The full chiral superfields obey
\be
\bar D_{\dot\a}Z=0~,~D_\a\bar Z=0~,
\ee
which leads us to define SIM-superspace superfields
\be\label{sz}
Z~,~\psi_-\equiv q_-Z~,~\bar Z~,~\bar\psi_\dm\equiv \bar q_\dm\bar Z~;
\ee
these obey the SIM-superspace constraints
\be\label{sc}
\bar d_\dpl Z=d_+\bar Z=0~,~
\bar d_\dpl \psi_-=i\pa_{-\dpl}Z~,~d_+\bar \psi_\dm=i\pa_{+\dm}\bar Z~.
\ee
The Lorentz symmetry-breaking term is
\be\label{2boun}
S_{SIM}=-im^2\int d_+\bar d_\dpl
\left(Z\frac1{\pa_{+\dpl}}\bar Z\right)~.
\ee
To go to components, we define 
$Z, \chi_+\equiv d_+ Z, \psi_-,F\equiv d_+\psi_-$
and the complex conjugates (actually, 
as follows from (\ref{dqdef},\ref{sz}), it makes sense
to identify $\chi_+\equiv\psi_+$, which we do below).
Using the constraints on the SIM-superfields
$Z,\psi$ and the algebra of spinor derivatives,
one recovers the component action:
\be\label{2comp}
S_{SIM}=m^2\int \left(Z\bar Z-i\,\psi_
+\frac1{\pa_{+\dpl}}\bar \psi_\dpl\right)~;
\ee
the Lorentz invariant terms are of course unchanged. A general
SIM-superspace action uses the measure $\int d_+\bar d_\dpl$; the 
Lagrangian is constructed out of the SIM-superspace superfields 
$Z,\psi_-,\bar Z,\bar\psi_\dm$ (\ref{sz},\ref{sc}) in such a way
that the net weight of $+$ and $\dpl$ indicies is minus one and
the $-,\dm$ indices enter only in scalar combinations.
Thus, for example, we find novel terms such 
as\footnote{This is SIM(2) invariant if $d_+Z$ and $\psi_-$ transform
as a the components of a single Weyl spinor, 
as follows from (\ref{dqdef},\ref{sz}).}
\be
S_{new}=-im\int d_+\bar d_\dpl
\left(d_+Z\frac1{\pa_{+\dpl}}\psi_-\right)+ c.c.~,
\ee
which gives rise to
\be
S_{new}=m\int 
\left(\psi_+\frac{\pa_{-\dpl}}{\pa_{+\dpl}}\psi_+
-ZF-\psi_+\psi_-\right)+ c.c.~.
\ee
We have not studied the physical consequence of these 
novel mass terms, but note that they resemble Majorana
masses and thus cannot arise for charged fields\footnote{For
completeness, we give their four-component form;
the fermionic terms are:
$ 
\psi_+\frac{\pa_{-\dpl}}{\pa_{+\dpl}}\psi_+
-\psi_+\psi_-=\psi_+\frac1{\pa_{+\dpl}}(\pa_{-\dpl}\psi_+
-\pa_{+\dpl}\psi_-)
~\propto ~\bar\psi\frac{\ns\ds}{n\cdot\pa}\psi_L$.}.
\section{Local forms of the SIM(2) action}
One may introduce unconstrained complex auxiliary SIM-superfields $X^{+\dpl}$ 
to remove the nonlocality as follows:
\be\label{local}
S_{local}=im^2\int d_+\bar d_\dpl \left( X^{+\dpl}\pa_{+\dpl} \bar X^{+\dpl} 
+ X^{+\dpl}Z+\bar X^{+\dpl}\bar Z\right);
\ee
Upon integrating out $X^{+\dpl}$, this gives the 
nonlocal action (\ref{2boun}).
In four-component notation, this becomes 
\be\label{4local}
S_{local}=-m^2\int d^2\q_{SIM}\left( Xn\cdot\pa \bar X + XZ+\bar X\bar Z\right),
\ee
where $d^2\q_{SIM}$ is the SIM-superspace measure 
$\bar D\ns D\propto (\bar\z D_L)( \bar\z D_R)$, and $X$ transforms
so as to ensure the SIM(2) invariance of the action.

\section{Coupling to gauge multiplets}
\setcounter{equation}{0}
The coupling to gauge multiplets is completely straightforward.
The Lorentz invariant gauge multiplet action is unchanged; we
couple to matter fields by making the replacement
\be
Z\frac1{n\cdot\pa}\bar Z \to Z\frac1{n\cdot\na}\bar Z 
\ee
in the SIM(2) part of the action. Then we have
\beqs\label{gauge}
S_{SIM}&=&-im^2\int d_+\bar d_\dpl
\left(Z\frac1{\na_{+\dpl}}\bar Z\right)\nonumber\\[2mm] 
&\equiv&-im^2\int\na_+\bar\na_\dpl
\left(Z\frac1{\na_{+\dpl}}\bar Z\right)\nonumber\\[2mm] 
&=&m^2\int \left(Z\bar Z-i\,\psi_
+\frac1{\na_{+\dpl}}\bar \psi_\dpl\right)~.
\eeqs

An interesting question arises whether there are SIM(2)
Lorentz-noninvariant terms possible for the gauge fields
themselves. The answer, surprisingly, appears to be yes.
We define SIM(2) superfields $W_\a,f_{+-}-iD'
\equiv\na_-W_+,f_{--}\equiv \na_-W_-$ 
and their complex conjugates; here 
$f_{\a\b}\propto F^+_{\mu\nu}\g^{\mu\nu}$ is the
self-dual part of the gauge field strength in two-component
notation. Then we may write down terms such as
\be
S_{nonlocal}=-m^2\int d_+\bar d_\dpl
\left(W_+\frac1{\na^2_{+\dpl}}\bar W_\dpl\right)~.
\ee
Defining components $\l_\pm\equiv W_\pm, f_++\equiv \na_+W_+$ 
and the $\q$-independent components of $D',f_{+-}=f_{-+},f_{--}$,
we find
\be\label{2f}
S_{nonlocal}=m^2\int 
\left(-i\l_+\frac1{\na_{+\dpl}}\bar \l_\dpl
+f_{++}\frac1{\na^2_{+\dpl}}\bar f_{\dpl\dpl}
\right)~.
\ee

In four component notation, these become
\be
S_{nonlocal}=m^2\int (\bar d\ns d_L) 
\left(\bar W \ns\frac1{(n\cdot\na)^2}W\right)~,
\ee
and
\be\label{4f}
S_{nonlocal}=m^2\int 
\left(\bar\l \ns\frac1{n\cdot\na}\l_L-
n^\mu n^\nu F_{\mu\rho}\frac1{(n\cdot\na)^2}F_\nu{}^\rho\right)~,
\ee
respectively. A remarkable simplification occurs when we make the 
lightcone gauge choice $n\cdot A=0$: the entire nonlocality drops out
of the gauge fields, and the second term in nonlocal action (\ref{2f},\ref{4f})
reduces to the usual mass term $A^2$, supplemented with the gauge
condition\footnote{Of course, the usual
massive vector field is not gauge invariant, and is not
constrained by this condition, which drastically 
changes the physics} $n\cdot A=0$. This makes it seem unlikely that theory
can be renormalizable; nevertheless, it is surprising
that one can write down a gauge invariant mass term in four dimensions,
albeit one that is only SIM(2) invariant.

One may wonder if the gauge system truly violates Lorentz symmetry;
for the fermionic term, the stress-tensor
has nonsymmetric terms, and thus does not generate Lorentz transformations
\cite{cf}. One can do a similar calculation here, but there is a simpler argument:
Lorentz invariance implies that massive vectors have three degrees of freedom;
the massive vectors constructed here have only two degrees of 
freedom\footnote{As for massless fields, the gauge condition
$n\cdot A=0$ eliminates one degree of freedom, and the field $\nt\,\cdot A$
becomes a nondynamical Lagrange multiplier; 
since all representations of SIM(2) are one-dimensional, 
there is no reason to expect the usual (Lorentz-invariant) 
counting of states for massive fields.}.
Unfortunately, the same argument 
almost certainly implies that this mass term is not a phenomenologically viable
alternative to the Higgs mechanism. In particular, the equivalence 
theorem\footnote{We thank George Sterman for pointing
out the relevance of this theorem.} \cite{equiv} implies that 
in the light-cone gauge, the longitudinal modes of the gauge bosons
can be described by the couplings of scalar goldstone fields, and such
fields are not present when one introduces a mass for the vector bosons
by a Lorentz noninvariant but SIM-invariant term such as (\ref{4f}).

All the work in \cite{cg,cg1,cf}, as well as ours, implicitly assumes
that $n\cdot\pa\ne0$; as emphasized to us by Erik Verlinde, this
implies a number of subtleties--in particular, modes annihilated by
$n\cdot\pa$ do not become massive, and in the gauge $n\cdot A=0$,
the mass term is not quite $A^2$, but rather has an implicit projection to
leave a residual gauge-invariance under gauge transformation with
parameters $\o$ obeying $n\cdot\pa\o=0$.

\vspace{10mm}
\noindent{\bf{Acknowledgement:}}
\noindent{MR thanks Dan Freedman for encouragement, 
discussions, comments, and
for reading several drafts of this manuscript, Kostas
Skenderis and Erik Verlinde for comments and discussions, 
George Sterman for discussions,
and the Institute for Theoretical Physics
at the University of Amsterdam for hospitality.
The work of UL is supported in part by 
VR grant 621-2003-3454 and by 
EU grant (Superstring theory) MRTN-2004-512194.
MR is supported in part by
NSF grant no.~PHY-0354776, 
by the University of Amsterdam, and by  
Stichting FOM.}


\end{document}